\documentclass[a4paper,useAMS,usenatbib]{mnras}
\usepackage{graphics,epsfig,psfig}
\usepackage[normalem]{ulem}
\usepackage{xcolor}
\usepackage{float}
\usepackage{subfig}
\usepackage{graphicx}
\usepackage[]{inputenc,amssymb}
\usepackage{widetext}
\usepackage{amsmath}
\usepackage[style=base, singlelinecheck=off, font=small, labelfont=bf, justification=raggedright]{caption}
\def \be{\begin{equation}}
\def \ee{\end{equation}}
\def \bea{\begin{eqnarray}}
\def \eea{\end{eqnarray}}

\def\prd{PRD}
\def\aap{A\&A}
\definecolor{webgreen}{rgb}{0,.5,0}
\definecolor{webbrown}{rgb}{.6,0,0}

\def\aap{A\&A}

\def\apjs{ApJS}
\def\apjl{ApJL}

\def\prd{Phys.Rev.D}         

\title[Astrophysical stochastic gravitational wave background]{Time-dependence of the  astrophysical stochastic gravitational wave background}
\author[Mukherjee\& Silk]{Suvodip Mukherjee$^{1,2}$\thanks{mukherje@iap.fr} \& Joseph Silk$^{1,2,3,4}$ \thanks{silk@iap.fr}\\
$^{1}$ Institut d'Astrophysique de Paris,  98bis Boulevard Arago, 75014 Paris, France\\
$^{2}$ Sorbonne Universites, Institut Lagrange de Paris,  98 bis Boulevard Arago, 75014 Paris, France\\
$^{3}$ The Johns Hopkins University, Department of Physics \& Astronomy, 3400 N. Charles Street, Baltimore, MD 21218, USA\\
$^{4}$ Beecroft Institute for Cosmology and Particle Astrophysics, University of Oxford, Keble Road, Oxford OX1 3RH, UK\\
}
\pubyear{2019}
\begin{document}
\label{firstpage}
\pagerange{\pageref{firstpage}--\pageref{lastpage}}
\maketitle
\begin{abstract}
The astrophysical stochastic  gravitational wave background (SGWB) is mostly produced from unresolved stellar binary mergers, and the number of events at any moment of time is expected to be  Poisson-distributed. The event rate is governed by several astrophysical processes. The Poisson nature leads to variation in the number of sources and this  causes temporal variations in the SGWB. The intrinsic temporal fluctuations of the SGWB are a rich source of astrophysical information that can be explored via ongoing and future gravitational wave experiments to classify the sources of the SGWB signal.  Along with several other methods to estimate the GW event rates from individual  sources, the study of the temporal variations of the SGWB signal provides an independent method for estimating the event rates of the GW sources that contribute to the  SGWB. Along with direct estimates of event rates, this approach can also distinguish between different sources contributing to the SGWB signal and will  be a useful probe of its evolution over a vast cosmic volume. On averaging over  observation times, the SGWB will be statistically invariant under time translation. Statistical time translation symmetry of the SGWB is expected due to the negligible evolution of the relevant cosmological and astrophysical phenomena over the observation time-scales over which the data is collected.
\end{abstract}

\section{Introduction}
Gravitational wave (GW) \citep{Einstein:1918btx} are a new multi-frequency observational probe of the Universe which will  explore a vast range of cosmological redshifts. The GW signal can be produced from astrophysical sources such as black holes (BHs), neutron stars (NSs), white dwarfs (WDs), supernovae  \citep{Vishveshwara:1970zz,Press:1971wr,Chandrasekhar:1975zza, Blanchet:1995ez, Buonanno:1998gg, Damour:2001bu, Blanchet:2004ek, Baker:2005vv,Pretorius:2005gq,Campanelli:2005dd, Buonanno:2004tp,Marassi:2011si, Zhu:2010af,Cusin:2018rsq}. Along with the astrophysical origins of GWs, it can also be generated in different cosmological scenarios such as during the period of inflation  \citep{Starobinsky:1979ty,Turner:1996ck, Martin:2013nzq}, cosmic strings  \citep{Kibble:1976sj,Damour:2004kw}, phase transitions \citep{Kosowsky:1992rz,Kamionkowski:1993fg} etc. The observed GW signal can be classified as point source or stochastic in nature. The GW signal can be resolved individually in the former case, and will be an unresolved continuous/discontinuous diffused background for the latter case.

The stochastic gravitational wave background (SGWB) \citep{Allen:1996vm, Maggiore:1999vm, Wu:2011ac,Regimbau:2007ed,Romano:2016dpx,Rosado:2011kv,Zhu:2011bd} spans a wide frequency range and is expected to have contributions from both astrophysical binary astrophysical sources (BHs, NSs, WDs, BH-NS, NS-WD) as well as from the cosmological origins.
One of the potential candidates of the cosmological SGWB is the primordial GW background produced during the epoch of inflation. This is can be explored at the low frequency ($f \sim 10^{-18}$ Hz) using the large angular scale B-mode polarization of the cosmic microwave background (CMB) which is accessible from  CMB experiments such as BICEP-KECK \citep{Ade:2018iql}, Simons Observatory\citep{Ade:2018sbj}, and LiteBIRD \citep{2018JLTP..193.1048S}. The high-frequency  primordial inflationary GW signal can in principle be directly probed if the astrophysical SGWB can be successfully  {distinguished from the inflationary GW signal}.  

In this work, we focus on the astrophysical SGWB which is produced from a large number of binary mergers of compact objects \citep{Rosado:2011kv, Regimbau:2008nj,Regimbau:2007ed, Wu:2011ac}. Such sources will contribute to the total GW energy background $\rho_{GW}$, which can be written in terms of the dimensionless quantity 
\begin{eqnarray}\label{eqgw1}
\Omega_{GW}(f)= \frac{1}{\rho_cc^2} \frac{d \rho_{GW} (f)}{d\ln f},
\end{eqnarray}
where $c$ is the speed of light  and $\rho_c= 3H_0^2/8\pi G$ is the critical density of the Universe, which depends on the present value of the Hubble constant $H_0$.  The astrophysical SGWB is expected to be anisotropic and several previous methods \citep{Mitra:2007mc,Thrane:2009fp, Talukder:2010yd, Mingarelli:2013dsa, Romano:2016dpx} were developed to measure the signal from the GW data. The currently ongoing ground-based GW experiments (such as  advanced Laser Interferometer Gravitational-Wave Observatory (LIGO) (Hanford and Livingston)  \citep{TheLIGOScientific:2014jea},  advanced VIRGO detectors  \citep{TheVirgo:2014hva}) and the upcoming ground-based GW experiments such as KAGRA \citep{Akutsu:2018axf} and  LIGO-India \citep{Unnikrishnan:2013qwa} are going to be operational in the coming decade to measure  {the GW signal in the frequency range $f \approx 30-3000$ Hz with a strain noise $\sim 10^{-23}\, {Hz^{-1/2}}$.} The data from the advanced-LIGO's first observational run have imposed upper bounds on both spatially varying and non-varying contributions to $\Omega_{GW}(f)$ \citep{TheLIGOScientific:2016dpb,TheLIGOScientific:2016xzw}.  {Another window to the GW signal is through} the Pulsar Timing Array (PTA) \citep{2010CQGra..27h4013H}, which are looking for GW signals in the frequency band $10^{-9}- 10^{-6}$ Hz and have imposed  constraints on the strain of SGWB signal as $1.45 \times 10^{-15}$ at $f=1\, \text{yr}^{-1}$  \citep{Arzoumanian:2018saf}. In the future, space-based GW observatory Laser Interferometer Space Antenna (LISA) \citep{2017arXiv170200786A} will  probe the frequency band $f \sim 10^{-5}-10^{-1}$ Hz of GW signals.   {In even longer timescale, the third generation GW experiments \citep{Evans:2016mbw} such as Einstein Telescope and Cosmic Explorer are going to measure GW signal for frequencies above $10$ Hz with sensitivity better than $10^{-24}\, {Hz^{-1/2}}$ \citep{Evans:2016mbw}. The third generation detectors will be able to measure the GW sources up to a redshift $z \sim 20$ with an snr $\sim 10$ \citep{Evans:2016mbw}.}

In this paper, we discuss the origin of temporal dependence in the astrophysical SGWB signal and  {show how this can be used to  distinguish between the SGWB signal originating from astrophysical and cosmological sources.} The temporal dependence is useful for probing the astrophysical event rates of the SGWB sources, the variations  with frequency of GW emission, and the spatial positions of the sources.  {The study of temporal fluctuations along with  spatial anisotropies bring a new dimension for exploring the SGWB background and its statistical properties. We show that this avenue is going  be useful for observationally distinguishing between the cosmological and astrophysical SGWB.} In Sec. \ref{time}, we discuss the origin of the temporal dependence of the SGWB signal. In Sec. \ref{aspects} and Sec. \ref{formalism}, we discuss the formalism and the corresponding estimator for studying the frequency and temporal fluctuations of the SGWB signal. The measurability of the time variability for a network of detectors such as advanced-LIGO (Hanford and Livingston), Virgo detectors and Cosmic Explorer are shown in Sec. \ref{error-estimate}. The conclusion and future scope of this work are discussed in Sec. \ref{conclusion}.

\section{Origin of the temporal dependence in SGWB} \label{time}
Along the cosmic light-cone in a particular sky direction and over a particular observational time window $\Delta t$, the SGWB signal has contributions from all of the events coalescing along the line of sight. The number of coalescing events taking place at different times is governed by  Poisson statistics as each of the events happens independently \footnote{It is a reasonable assumption to consider that one coalescing binary system is not triggering other binary events.}. The corresponding probability mass function of occurrence of $N$ mergers of binaries of mass $M$  in a time interval $\Delta t$ can be written as
\begin{equation}\label{poisson}
P (N, M)= \frac{e^{-\Lambda (M,\dot \rho, z)\Delta t} (\Lambda (M,\dot \rho, z) \Delta t)^{N}}{N!},
\end{equation}
where $\Lambda (M,\dot \rho,z)$ is the average event rate \citep{Kalogera:2006uj, OShaughnessy:2006uzj, Bulik:2008ab,Mandel:2009nx,Dvorkin:2017kfg} and is still poorly known from astrophysical observations and is expected to depend on several parameters such as the mass of the binary compact objects $M$, star formation history $\dot \rho$, the source redshift $z$. In the above equation the time in the observer's frame $\Delta t$ is related to the time in the source rest frame $\Delta t_r$ at redshift $z$ by the relation $\Delta t= \Delta t_r (1+z)$. 

The Poisson nature of the GW events results in a variation in the number of GW sources at different time intervals $\Delta t$ with the standard deviation proportional to $(\Lambda (M,\dot \rho, z) \Delta t)^{1/2}$. As a result, the SGWB, which is an integrated effect of all of the events, will exhibit a temporal variation at a fixed direction in the sky.  The average value of the event rate is expected to be constant as it is governed by the astrophysical and cosmological phenomena, and as a result, we can expect negligible evolution of the event rate $\Lambda (M, \dot \rho, z)$ over the observation time. This implies that the SGWB will be \textbf{time-dependent}, but will exhibit \textbf{statistical time translation symmetry} on averaging over large observation time. For large event rates, the Poisson distribution will tend towards a Gaussian distribution (central limit theorem), as the skewness and kurtosis decreases as $(\Lambda(M,\dot \rho, z) \Delta t)^{-1/2}$ and $(\Lambda(M,\dot \rho, z) \Delta t)^{-1}$ respectively. However the variance of the signal grows with the event rates according to the relation $\Lambda(M,\dot \rho, z) \Delta t$ indicating that the rms time variability of the SGWB signal will not disappear due to the large event rate. 

We can write the observed $\Omega_{GW} (f)$ at a frequency $f$ in terms of the energy emission from each GW source and number of GW events  $N(z)$ in a comoving volume between the cosmic time $t(z)$ and $t(z+\Delta z)$ as  \citep{2001astro.ph..8028P}
\begin{align}\label{basic-1}
    \begin{split}
        \Omega_{GW} (f)= \frac{1}{\rho_cc^2}\int dz \frac{N(z)}{(1+z)} \frac{dE_{GW}}{d\ln f_r}\bigg|_{f_r= (1+z)f}, 
    \end{split}
\end{align}
where, $\rho_c= 3H^2_0/ 8\pi G$ and $c$ is the speed of light. The variation in the amplitude of the SGWB signal within the time interval $\Delta t$ will be reflected due to the change in the number of GW events $\Delta N (z, \Delta t)/ N(z, t) = N(z, t+\Delta t)/N(z, t) -1$ in the time interval $\Delta t$. 
Due to Poisson nature of the GW events, if the variation in the number of sources $\Delta N (z, \Delta t)$ (within a time interval $\Delta t$) is comparable to the total number of GW sources ($N (z, t)$) contributing to the SGWB signal at frequency $f$ and time $t$, then the relative fluctuations in the SGWB signal around the mean value of $\Omega_{GW} (f)$ are of the order $\approx \Delta N(z, \Delta t)/ N(z, t)$. This implies that the maximum time variability of the SGWB signal is going to be evident when the condition $\Delta N (z) \sim N (z)$ will be satisfied. 

The time variability of the SGWB signal depends primarily on three time-scales, (i) the duration $\tau_d$ which a GW source spends at a particular frequency $f$, (ii) the duration between the consecutive events ($\Delta t_{event} \propto 1/\Lambda$), and (iii) the time-scale $\Delta t$ over which we  estimate the variation of the sky signal. For the first two time-scales $\tau_d$ and $\Delta t_{event}$, we can define the duty cycle of the GW signal as the ratio of the duration of the signal emitted between  frequency $f$ and $f+ \Delta f$, and the time difference between two GW events \citep{Wu:2011ac, Rosado:2011kv}  
\begin{align}\label{basic-2}
    \begin{split}
\frac{d\mathcal{D}}{df}= \int dz \, \dot n(z) \frac{d\tau_d}{df},    \end{split}
\end{align}
where $\dot n (z)$ is the event rate as a function of the comsological redshift and the duration of the signal $\frac{d\tau_d}{df}$ at frequency $f$  can be written in terms of the GW chirp mass according to the relation
\begin{eqnarray}\label{eqgw7}
\frac{d\tau_d}{df}= \frac{5c^5}{96\pi^{8/3} G^{5/3} \mathcal{M}_z^{5/3} f^{11/3}}.
\end{eqnarray}

A large duty-cycle implies that the duration of the  GW signal at  frequency $f$ is long compared with the time difference between the events. So, if $\Delta t < \Delta t_{event}$, the sources of the GW signal are not changing within the time $\Delta t$, resulting in a SGWB signal with no temporal fluctuations. Also, when $\Delta t < \tau_d$, the variations of the GW signal at a frequency $f$ is also negligible within the time scale $\Delta t$. This implies for the scenario $\tau_d >> \Delta t_{event} >> \Delta t$, the number GW sources produced in the time $\Delta t$ is much smaller than the total number of GW sources contributing to the SGWB signal and also the intrinsic time variability of the signal in the time scale $\Delta t$ is negligible. As a result, the fractional change in the SGWB signal with respect to the average background is tiny for this case. 

On the other hand, if the three time-scales satisfies the condition ($\Delta t \geq \Delta t_{event} \geq \tau_d$), then the SGWB  signal shows time variability comparable to the mean average signal in the time scale $\Delta t$. The GW sources  satisfying this criterion are expected to contribute maximally to the time variability of the SGWB signal. For this kind of source, the duty cycle is of the order unity or less, i.e $\tau_d \leq \Delta t_{event}$. The condition for the time variability of the SGWB signal can also be written in terms of the overlap function as defined by \citet{Rosado:2011kv}. The contribution from the non-overlapping GW sources will cause dominant contribution to time variability in the SGWB signal. Whereas the overlapping sources for which $\Delta N (z) << N (z)$, is going to produce negligible time variability in the SGWB.

\begin{figure}
    \centering
    \includegraphics[trim={0 0 0 0cm}, clip, width=0.5\textwidth]{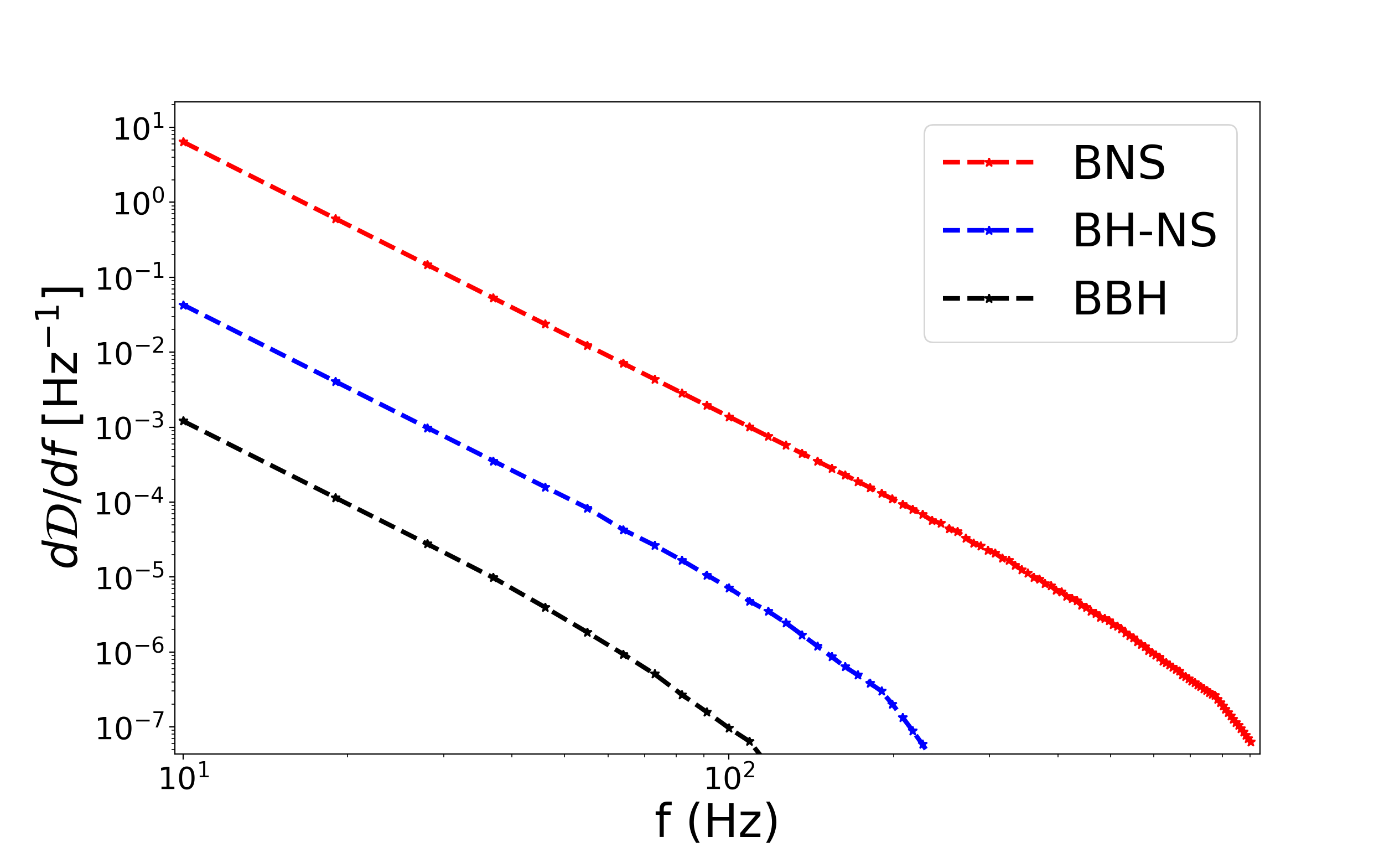}
    \caption{The duty-cycle for binary black holes (BBH), binary neutron stars (BNS) and black holes- neutron star (BH-NS) are shown for ground-based detectors.}
    \label{fig:duty}
\end{figure}

In Fig. \ref{fig:duty}, we plot the duty-cycle as a function of the GW frequency for binary neutron stars (BNS), black hole neutron star (BH-NS) and binary black holes (BBH). For this plot, we have taken the event rate of GW sources as $1$  Mpc$^{-3}$ Myr$^{-1}$, $0.03$  Mpc$^{-3}$ Myr$^{-1}$ and $0.005$  Mpc$^{-3}$ Myr$^{-1}$ for BNS, BH-NS and BBH respectively.  From Fig. \ref{fig:duty}, it is evident that BNS, followed by BH-NS systems, are going to have overlapping sources resulting in less temporal fluctuations in the SGWB signal as opposed to the BBH systems. As a result, this avenue helps in distinguishing between the SGWB originating from the BNS, BH-NS and BBHs systems.

\begin{figure}
    \centering
    \includegraphics[trim={0 0 0 0cm}, clip, width=0.5\textwidth]{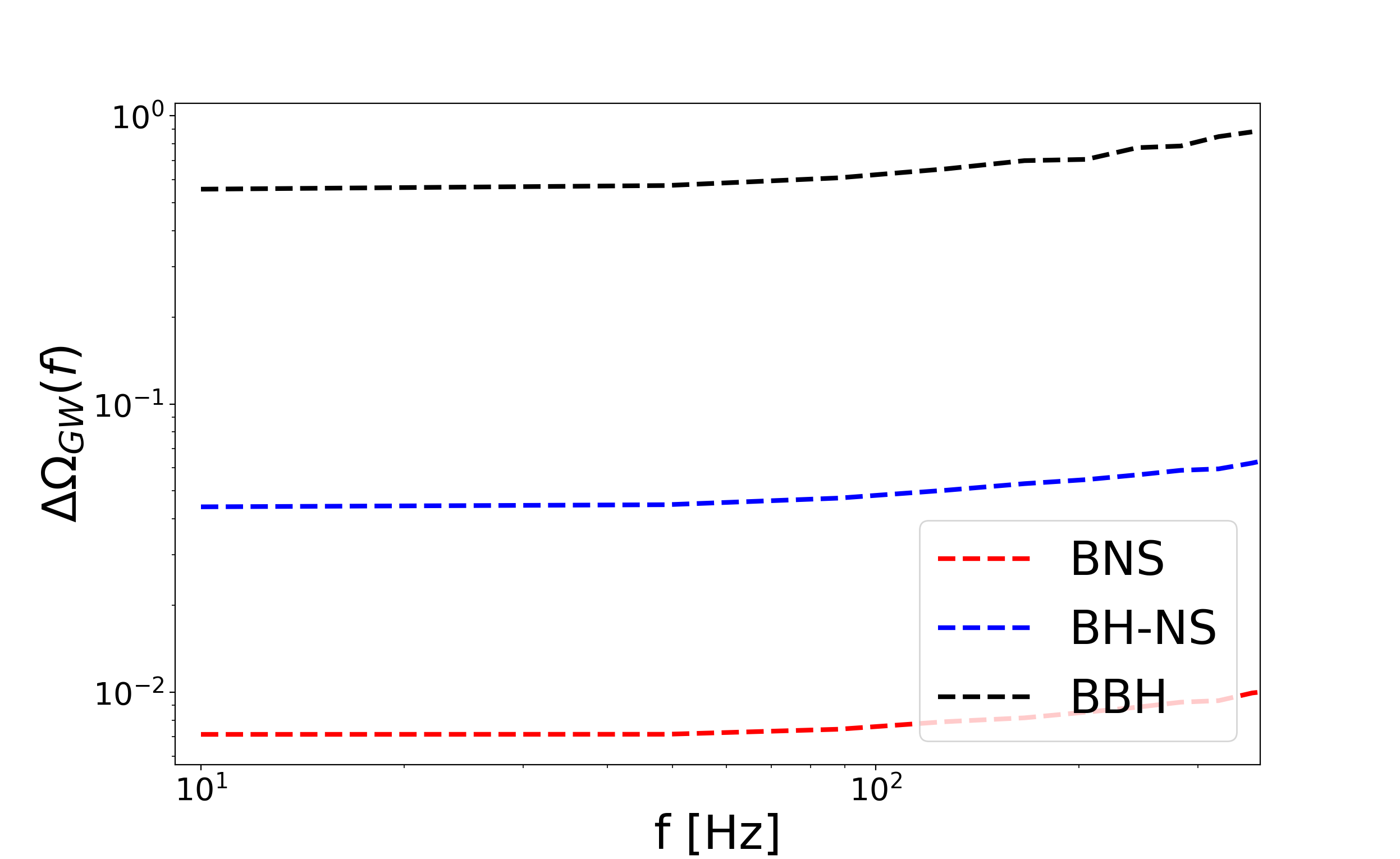}
    \caption{We show the rms fluctuations in the SGWB signal expected due to the Poisson nature of the GW binary sources. The rms fluctuations for three species of binaries namely binary black holes (BBH), binary neutron stars (BNS) and black holes- neutron star (BH-NS).}
    \label{fig:rms}
\end{figure}

\begin{figure}
    \centering
    \includegraphics[trim={0 0 0 0cm}, clip, width=0.5\textwidth]{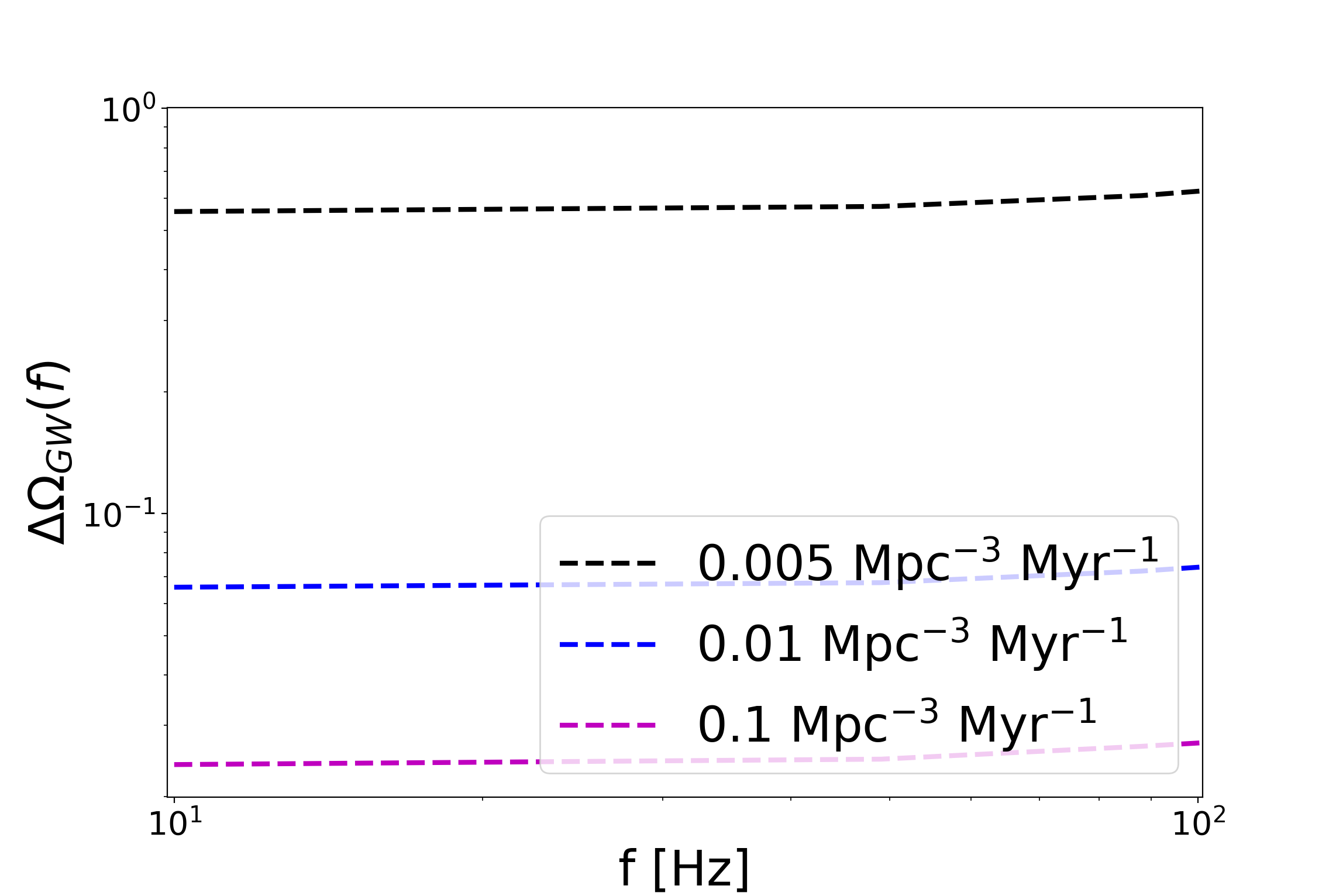}
    \caption{We show the rms fluctuations in the SGWB signal expected due to the Poisson nature of the GW binary black holes with three different rates. With  an increase in the number of events, the background rms fluctuations diminish for the frequency band $10-100$ Hz, which are explored by ground based GW detectors such as LIGO and VIRGO.}
    \label{fig:rmsbh}
\end{figure}

To show the time variability of the SGWB for BNS, BH-NS and BBH, we estimate  the rms temporal fluctuations of the SGWB\footnote{Here the notation $\langle .\rangle_T$ implies average in the time domain for different realizations of SGWB.} $\Delta \Omega_{GW} (f)= \sqrt{\bigg\langle \bigg(\frac{\Omega_{GW} (t,f) - \bar \Omega_{GW} (f)}{\bar \Omega_{GW} (f)}\bigg)^2\bigg\rangle_T}$ from $100$ realizations of $\Omega_{GW} (t)$ drawn from a Poisson distribution with the fixed event rates $1\, \text{Mpc}^{-3}\,\text{Myr}^{-1}$, $0.03\, \text{Mpc}^{-3}\,\text{Myr}^{-1}$ and $0.005\, \text{Mpc}^{-3}\,\text{Myr}^{-1}$ for BNS, BH-NS and BBH respectively. The corresponding rms variations in the SGWB signal are shown in Fig. \ref{fig:rms} for the frequency range which are relevant for the terrestrial GW observatories such as advanced-LIGO, VIRGO and KAGRA, LIGO-India. The maximum relative temporal fluctuations of the SGWB signal are expected from BBHs as the number of overlapping events is fewer, resulting from a smaller value of the duty cycle. On the other hand, BNS will exhibit less relative time variability due to their larger duty cycles. This behavior of the SGWB over these frequency ranges will allow us to distinguish between the different types of  GW sources which are contributing to the signal. The amplitude of the time variability of the SGWB signal depends strongly on the event rate, probability distribution of the compact object mass and their redshift distribution. We show the changes in the rms fluctuation for the BBH case with the variation in the event rates and probability distributions of the black hole masses in Fig. \ref{fig:rmsbh}.  Fig. \ref{fig:rms} and Fig. \ref{fig:rmsbh} indicates that  characterizing the time variability of the SGWB signal is a useful probe to understand different populations of binary GW source and their event rates respectively. 


 {The time variability of the SGWB signal is a  independent avenue along with other existing methods to detect the GW event rates from the individual GW sources and SGWB signal \citep{2013NJPh...15e3027M, 2015PhRvD..91b3005F, PhysRevX.8.021019}. These frameworks \citep{2013NJPh...15e3027M, 2015PhRvD..91b3005F, PhysRevX.8.021019} are able to measure the rates from the GW data and can successfully distinguish between the background and individual signal. The detection of the individual GW events are going to be more powerful in detecting the local GW event rates from the "loudest individual events". However, the study of the temporal variation of the SGWB signal will be powerful in mainly distinguishing the population of different SGWB sources and understanding their distribution over the cosmic volume. The event rates that can be inferred from the SGWB signal is capable to probe the signal up to a high redshift which cannot be inferred from the individual loud sources. The temporal dependence of the SGWB signal can also be a useful avenue for distinguishing between astrophysical SGWB signals and the cosmological SGWB signals, since the latter are generally not expected to show sporadic behavior in the temporal domain. 
}

The temporal fluctuations of  $\Omega_{GW}(f, \hat n, t)$ can be written as 
 \begin{eqnarray}\label{eqgw4}
\begin{split}
\Omega_{GW}(f,\hat n, t)=&  \frac{f}{4\pi\rho_c c^3} \iint dz\,d\mathcal{M} R_\mathcal{M}(z,\hat n, t)\frac{dV}{dz}\\ & \times \overbrace
{\bigg(\frac{(1+z)f_r}{d_L^2} \frac{dE_{GW}(f_r, \hat n)}{df_r}\bigg)}^{\text{GW source}},
\end{split}
\end{eqnarray}
where we can express the emission of the GW signal during the inspiral phase from an individual source in terms of the redshifted chirp mass $\mathcal{M}_z= \mathcal{M} (1+z)$ and redshifted frequency $f_r= (1+z)f$ by the relation
\begin{eqnarray}\label{eqgw4b}
\frac{dE_{GW}(f_r, \hat n)}{df_r}= \frac{(G\pi)^{2/3} \mathcal{M}_z^{5/3}}{3f_r^{1/3}}.
\end{eqnarray}
The number of coalescing binaries $R_\mathcal{M}(z,\hat n, t)$ per unit earth time, per unit comoving volume, per unit mass-bin for the binaries of redshifted chirp mass $\mathcal{M}$ can be written as
\begin{eqnarray}\label{eqgw4b}
R_\mathcal{M}(z,\hat n, t)= \frac{\overbrace{n_{gal}(\hat n, z)}^{\text{cosmology}}\overbrace{\mathcal{E}_\mathcal{M}(z,\hat n, t_r)}^{\text{astrophysics}}}{(1+z)},
\end{eqnarray}
where, $n_{gal}(\hat n, z)$ is the number density of galaxies and $\mathcal{E}_\mathcal{M}(z,\hat n, t_r)$ is the number of GW binaries coalescing per galaxy, per unit cosmic time, per unit chirp mass. $dV/dz$ denotes the comoving volume element which can be written as $4\pi c d_L^2/ ((1+z)^2H(z))$. The merger and the ring-down phase of the individual GW signal can also be included to model the SGWB signal. As the GW events are expected to be a statistical process (Poisson distributed), at any moment of time, $\mathcal{E}_\mathcal{M}(z,\hat n, t)$ is going to be time-dependent. 

The average of $\Omega_{GW} (f)$ over a large observation time will exhibit statistical time translation symmetry which can be written as 
\begin{eqnarray}\label{eqgw3}
\bar\Omega_{GW} (f)\equiv \langle \Omega_{GW}(f, \hat n, t) \rangle_T = \frac{1}{T}\int dt \frac{1}{\rho_c} \frac{d \rho_{GW} (f, \hat n, t)}{d\ln f}.
\end{eqnarray}

\section{Aspects of  SGWB in frequency and temporal domains}\label{aspects} 
We propose to study three aspects of the  astrophysical SGWB:  (i) \textit{Spectral derivative of the SGWB signal}, (ii) \textit{Time derivative of the SGWB signal} and, (iii) \textit{Tomographic redshift estimation of the SGWB signal}. 

\subsection{Spectral derivative of the GW signal at a fixed observation time}
The spectrum of the GW signal is an indicator of the mass of the coalescing binaries, as the mergers of the binaries with heavier masses happen at a lower frequency. We can then construct the spectral derivative of the SGWB signal  as
 \begin{align}\label{sgwbft}
\begin{split}
\mathcal{F} (t, f_a, f_b, \hat n)\equiv & \Omega(t, f_a, \hat n)- \bigg(\frac{f_a}{f_b}\bigg)^{2/3}\Omega(t, f_b, \hat n)  \\ =& \frac{(G\pi)^{2/3}f_a^{2/3}}{3\rho_c c^2 H_0}  \iint dz \, d\mathcal{M} \, \frac{n_{gal}(\hat n, z)}{(1+z)^{4/3}E(z)} \\ &\bigg[\bigg(\mathcal{E}_{\mathcal{M}}(z,\hat n, t_r) \mathcal{M}_z^{5/3}(t, \hat n)\bigg) \bigg{|}_{f_a}\\ & - \bigg(\mathcal{E}_{\mathcal{M}}(z,\hat n, t) \mathcal{M}_z^{5/3}(t, \hat n)\bigg)\bigg{|}_{f_b}\bigg],
\end{split}
\end{align}
where, $f_a= f_b+\Delta f$ and we have used the fact that the GW strains from the individual binaries are not correlated with other coalescing binaries. $E(z)= \sqrt{\Omega_m(1+z)^3 + \Omega_\Lambda}$ is the expansion history of the Universe. The maximum frequency ($f_{max}$) up to which the GW signal is emitted depends on the last stable orbit of the BBH, BH-NS, and BNS. This depends on the total mass $M$ of the coalescing binaries as $f_{max}=  \frac{c^3}{6\sqrt{6}\pi G M}$. So the GW signal emitted from sources with the mass of the coalescing binaries between $M^a\propto \frac{1}{f_a}$ and $M^b\propto \frac{1}{f_b}$
 will contribute to the differential signal $\mathcal{F} (t, f_a, f_b, \hat n)$ 
 
 \begin{align}\label{sgwbft2}
\begin{split}
\mathcal{F} (t, f_a, f_b, \hat n)=& \frac{(G\pi)^{2/3}f_a^{2/3}}{3\rho_c c^2 H_0} \int dz\, n_{gal}(\hat n, z)\\ & \times\int d\mathcal{M} \bigg(\frac{\mathcal{E}_{\mathcal{M}}(z,\hat n, t) \mathcal{M}_z^{5/3}(t, \hat n)}{(1+z)^{4/3}E(z)}\bigg) \\ &\times \bigg(H({M}^a_{z}-{{M}_{z}})-H({M}^b_{z}-{{M}_{z}})\bigg),
\end{split}
\end{align}
where $H(x)$ is the Heaviside function which is non-zero only for positive values of the argument $x$. For the same observation time and sky direction, the number of GW sources emitting in individual mass bins is fixed. As a result, the spectral derivative of the SGWB captures the sky luminosity in the redshift-averaged mass-bins corresponding to $\Delta f/f$. This will also exhibit temporal fluctuations due to the variation in the number of events with cosmological redshift at a fixed frequency bin-width. However, the temporal dependence can be weak, depending on the population of the sources and the event rate.
   
The spatial and time average of the spectral derivative signal in individual mass-bins can be expressed as 
 \begin{eqnarray}\label{sgwbft3}
\bar{\mathcal{F}} (f_a,f_b)= \frac{1}{4\pi T_{obs}}\int d^2\hat n \int_0^{{T}_{obs}} dt\, \mathcal{F} (t, f_a, f_b, \hat n).
\end{eqnarray}
$\bar{\mathcal{F}} (f_a,f_b)$ can be expected to be a constant over separate large observational windows due to the negligible variation of the astrophysical and cosmological phenomena over the observation time.  
\subsection{Temporal derivative of the SGWB signal at  fixed frequency}
The temporal dependence due to the Poisson distribution can be captured by estimating the variation in the SGWB amplitude between two epochs of time $t$ and $t'= t-\Delta t$ at a fixed GW frequency $f$ 
 \begin{align}\label{sgwbtime1}
\begin{split}
\mathcal{T} (t, t', f, \hat n) \equiv\,& \Omega (t, f, \hat n)- \Omega (t', f, \hat n)\\ =& \frac{(G\pi)^{2/3}f^{2/3}}{3 \rho_c c^2} \iint dz \, d\mathcal{M}\frac{n_{gal}(\hat n, z)}{(1+z)^{4/3}E(z)}\\& \times \bigg[\bigg({\mathcal{E}_{\mathcal{M}}(z,\hat n, t)\mathcal{M}_z^{5/3}(\hat n)}\bigg) \\ & - \bigg({\mathcal{E}_{\mathcal{M}}(z,\hat n, t')\mathcal{M}_z^{5/3}(\hat n)}\bigg)\bigg].
\end{split}
\end{align}

This is the line-of-sight integrated quantity of the variation of the number of events in any sky direction. The spectrum of this quantity is a direct probe of the variation of the event rate across the GW source population. 

\subsection{Tomographic redshift estimation of the spatial distribution of the GW sources}\label{position}
The three-dimensional spatial distribution of the GW sources depend upon the background cosmology, the redshift of the sources and the spatial distribution of the host galaxy. Cosmological information (such as the redshift distribution of the host galaxies of the GW sources)  which is embedded in the SGWB signal is not time-dependent. The contribution of the SGWB signal from individual cosmological redshifts can be obtained by cross-correlating  with redshift-known cosmic probes of the density field. From a galaxy survey, we obtain the spatial position of galaxy along with its photometric or spectroscopic redshift. We can construct the tomographic density field at different cosmological redshifts as
 \begin{eqnarray}\label{cosmo1}
\delta_{g} (\hat n, z)= \frac{n_{gal}(\hat n, z)}{\bar n_{gal}(z)}-1,
\end{eqnarray}
where $n_{gal}(\hat n, z)$ is the number of galaxies in the redshift bin $z$ and in the direction $\hat n$ and $\bar n_{gal}(z)$ is the average number of galaxies. The two-point correlation function of galaxies $\mathcal{W}_{gg}$ is a probe to the underlying dark matter correlation function $ \xi_{DM} (\chi, z)$ given by the relation  
 \begin{align}\label{gal1}
\mathcal{W}_{gg}(\theta, z) \equiv & \langle \delta(\hat n, z)  \delta(\hat n + \theta, z)\rangle\nonumber \\ =&\int d^2\hat n\, D^2(z)b^{2}_{g}(z)\xi_{DM}( \chi(z'), \chi(z), \theta), 
\end{align}
where the angular bracket $\langle.\rangle$ denotes the 
all-sky average, $D(z)$ is the growth function, $b_g(z) = \delta_g/\delta_{DM}$ is the galaxy bias with respect to the dark matter density field $\delta_{DM}$ and the correction function $\xi_{DM}( \chi(z'), \chi(z), \hat \theta)$ is related to the power spectrum of the dark matter distribution $P_{DM}(k)$ by
 \begin{eqnarray}\label{corr}
\xi_{DM}( \chi(z'), \chi(z), \hat \theta)= \frac{1}{2\pi^2} \int dk k^2 P_{DM}(k) j_0(k\chi),
\end{eqnarray}
 where $j_0(k\chi)$ is the spherical Bessel function and $\chi= \sqrt{\chi(z')^2 + \chi(z)^2 - 2\chi(z')\chi(z')\cos{\theta}}$.  

The cross-correlation of the SGWB with the tomographic redshift bins of the large-scale structure encodes the cosmological spatial distribution of the GW sources. The 
time-dependent $\mathcal{F} (t, f, f', \hat n)$ and $\mathcal{T} (t, t', f, \hat n)$ signal intrinsically depends on  the redshift distribution of the GW sources. So, the angular cross-correlation of $\mathcal{F} (t, f, f', \hat n)$ with the tomographic density field $\delta_g (\hat n, z)$ can be expressed as 
 \begin{align}\label{sgwbfco1}
\mathcal{W}_{\mathcal{F}g}(t, f, z, \theta) \equiv& \langle \mathcal{F} (t, f, \hat n)  \delta(\hat n + \theta, z)\rangle\nonumber \\ =&\int d^2\hat n\int dz'\, D(z')D(z)b_{g}(z)b_{\mathcal{F}}(t, f,z',\hat n)\nonumber \\& \times \xi_{DM}( \chi(z'), \chi(z), \theta), 
\end{align}
where $b_{\mathcal{F}}(t, f, z', \hat n) $ is the time-dependent GW bias for the spectral derivative which we can express in terms of the GW source properties by the relation
 \begin{eqnarray}\label{sgwbfco1a}
b_{\mathcal{F}}(t, f, z, \hat n)=&\frac{(G\pi)^{2/3}f_a^{2/3}}{3\rho_c c^2 H_0} \int d\mathcal{M} \bigg(\frac{\mathcal{E}_{\mathcal{M}}(z,\hat n, t) \mathcal{M}_z^{5/3}(t, \hat n)}{(1+z)^{4/3}E(z)}\bigg)\nonumber \\ &\times \bigg(H({M}^a_{z}-{{M}_{z}})-H({M}^b_{z}-{{M}_{z}})\bigg).
\end{eqnarray}

Similarly, the cross-correlation of $\mathcal{T} (t, t', f, \hat n)$ with the  galaxy distribution will be able to capture the time evolution of the binary events with redshift, which can be written as
 \begin{align}\label{sgwbfco2}
\mathcal{W}_{\mathcal{T}g}(t, t', f', z, \theta) \equiv& \langle \mathcal{T} (t, t', f, \hat n)  \delta(\hat n + \theta, z)\rangle\nonumber \\= & \int d^2\hat n\int dz'\, D(z)D(z')b_{g}(z)\nonumber  \\& \times b_{\mathcal{T}}(t,t', f',z', \hat n) \xi_{DM}( \chi(z'), \chi(z), \theta), 
\end{align}
where, we have defined $b_{\mathcal{T}}(t,t', f',z, \hat n)$ as
 \begin{align}\label{sgwbfco1c}
b_{\mathcal{T}}(t,t', f',z, \hat n)=&\frac{(G\pi)^{2/3}f^{2/3}}{3 \rho_c c^2} \int  \, d\mathcal{M}\frac{1}{(1+z)^{4/3}E(z)}\nonumber \\& \times \bigg[\bigg({\mathcal{E}_{\mathcal{M}}(z,\hat n, t)\mathcal{M}_z^{5/3}(\hat n)}\bigg) \nonumber \\ & - \bigg({\mathcal{E}_{\mathcal{M}}(z,\hat n, t')\mathcal{M}_z^{5/3}(\hat n)}\bigg)\bigg].
\end{align}

The two-point cross-correlation $\mathcal{W}_{\mathcal{T}g}(t, t', f', z, \theta)$ and $\mathcal{W}_{\mathcal{F}g}(t, f, z, \theta)$ are all-sky integrated quantities. For a  statistically isotropic Universe, the all sky-average of the cross-correlation signal will translate into a temporal average over different realizations of the events. As a result, we can expect the two-point correlation function to have statistically time translation symmetry. It will be related to the mean merger rate of GW sources per unit redshift per unit time which is denoted by $\mathcal{E}_{\mathcal{M}}(z)$. 

The two-point cross-correlation function, defined in  real space, can also be obtained in the spherical harmonic basis of the field ($X_{lm}= \int d^2\hat{n} Y^*_{lm} (\hat n) X(\hat n)$) using the addition theorem of spherical harmonics
 \begin{eqnarray}\label{sph-func}
\mathcal{W_\mathcal{XX'}}(\theta)= \sum_l \bigg(\frac{2l+1}{4\pi}\bigg) P_l(\cos(\theta)) C^{XX'}_l,
\end{eqnarray}
where $C_l^{XX'}= \sum_m X_{lm}X'^*_{l'm'} \delta_{ll'}\delta_{mm'}$ and $P_l(\cos(\theta))$ are the Legendre Polynomials. 

In summary, the time dependence of the SGWB carries a rich source of information about the astrophysical event rate of the GW sources. We show that a set of observables such as $\mathcal{F}$, $\mathcal{T}$, $\mathcal{W}_{\mathcal{F}g}$, and $\mathcal{W}_{\mathcal{T}g}$ are capable of probing the event rate as a function of cosmological redshift and the chirp mass of the binaries from the SGWB signal. The auto-power spectrum of the SGWB is also an interesting probe. Recent studies has explored the auto-power spectrum and the impact of shot-noise \citep{Jenkins:2019uzp, Jenkins:2019nks}. A recent study \citep{Canas-Herrera:2019npr} has explored cross-correlation of the SGWB signal and galaxy surveys for a cosmic variance limited measurement of the SGWB and without considering the overlap reduction function of GW detectors.

\section{Estimators for the astrophysical  SGWB}\label{formalism}
In this section, we devise the formalism which will be useful for extracting the astrophysical time-dependence of the SGWB signal from the GW data. The extraction of the stochastic GW signal can be devised in both pixel-space (real space) and spherical harmonic space. We show the estimators which can be used to study the quantities such as $\mathcal{F}$, $\mathcal{T}$, $\mathcal{W}_{\mathcal{F}g}$, and $\mathcal{W}_{\mathcal{T}g}$.  

\subsection{Overview of the analysis technique}
In this section, we briefly explain the standard framework for the analysis of stochastic GW data \citep{Mitra:2007mc, Thrane:2009fp, Talukder:2010yd, Romano:2016dpx}. 
The time-series data $d_i(t)$ from the detector $i$ can be written as
 \begin{eqnarray}\label{es1}
d_i(t)= h_i(t) + n_i(t),
\end{eqnarray}
where, $h_i(t)$ and $n_i(t)$ are the signal and noise respectively. The observed signal can be written in terms of the true sky signal $h_p^s$ and the detector response $F^p_i (\hat n, t)$ as \citep{Romano:2016dpx, Thrane:2009fp}
 \begin{eqnarray}\label{es1a}
h_i(t) = \int_{-\infty}^{\infty} df\int d^2\hat n F^p_i (\hat n, t)e^{2\pi i f(t- \vec x_I. \hat n/c)}h^s_{p} (f, \hat n).
\end{eqnarray}
The short-time Fourier transform of the sky signal at a particular time (t) can be written as 
\begin{eqnarray}\label{ft}
d(t, f)=& \int_{t- \tau/2}^{t+\tau/2} d(t') e^{-2\pi ift'} dt',
\end{eqnarray}
where the choice of $\tau$ is made such that the detector response function has not changed significantly in the time window $\tau$. It should also not be smaller than the travel time of the GW signal between a pair of detectors. The expectation value of the cross-correlation of $d_i(f)$ between two different detectors is
 \begin{align}\label{es2}
\langle \Omega^d_{ij}(f, t) \rangle \equiv \frac{2}{\tau}\langle{d_i(f,t)d^*_j(f,t)}\rangle
=   \frac{2}{\tau}h_i(f, t)h^*_j(f,t),
\end{align}
where the noise terms between the detectors are expected to be uncorrelated $\langle{n_i(f)n^*_j(f)}\rangle=0$. The corresponding noise covariance matrix of $\Omega^d_{ij}(f, t)$ is 
 \begin{eqnarray}\label{es2cov}
 \mathcal{N}^d_{ij}(f, t, f',t') \equiv &\langle\Omega^d_{ij}(f, t)\Omega^d_{ij}(f', t') \rangle - \langle\Omega^d_{ij}(f, t)\rangle\langle\Omega^d_{ij}(f', t')\rangle,\nonumber \\
=& \mathcal{N}_i (f)\mathcal{N}_j (f) \delta_{tt'}\delta_{ff'}
\end{eqnarray}
where $\mathcal{N}_x$ is the one-sided noise power spectrum for detector $x$ which is uncorrelated with the noise of the other detectors. In this expression, the contribution of the signal is assumed to be negligible in comparison to the detector noise. 

With knowledge of the detector response function $F^p_i (\hat n, t)$ for both the detectors, we can relate the observed cross-correlation spectrum with the true signal in the sky $\Omega^s(\hat n, f, t)$ as 
 \begin{eqnarray}\label{es3}
\Omega^d_{ij}(f,t)= \int d^2 \hat n \gamma(\hat n, f, t)\Omega^s(\hat n, f, t),
\end{eqnarray}
where, $\gamma(\hat n, f, t)$ is the geometric factor  \citep{PhysRevD.48.2389, PhysRevD.46.5250}
 \begin{eqnarray}\label{es3a}
\gamma(\hat n, f, t)= \frac{1}{2}F^p_i (\hat n, t)F^p_j (\hat n, t)e^{2\pi i f\hat n.(\vec x_i- \vec x_j)/c}.
\end{eqnarray}

In Eq. \ref{es3}, we differ from the previous method in two ways. Firstly, the SGWB signal has an intrinsic temporal dependence $\Omega^s(\hat n, f, t)$. Secondly, we have not separated the frequency and the angular parts, but rather kept them together in $\Omega^s(\hat n, f, t)$. In the previous analysis \citep{Thrane:2009fp}, the temporal dependence of the data was considered to be only due to the temporal dependence of the geometric factor $\gamma(\hat n, f, t)$ and the sky signal was assumed to depend only on the sky direction. 

\subsection{Estimators}

\textbf{Spectral-derivative: } The spectral-derivative of the SGWB signal is the difference of the SGWB signal between two frequency channels and at a fixed detector time. The corresponding estimator in terms of the  cross-power spectrum of the GW data can be written as 
 \begin{align}\label{es5}
\Delta \hat \Omega_{\mathcal{F}} \equiv\, &  \Omega^d(f, t)-  \Omega^d(f', t), \nonumber \\=& \int d^2 \hat n \gamma(\hat n, f, t)(\Omega^s(\hat n, f, t)- \Omega^d(\hat n, f', t)),
 \end{align}
 where the frequency bands need to satisfy $|f-f'|= \Delta f << c/(2\pi D)$, such that the geometric factor does not change significantly in this narrow frequency band and with the precise knowledge of the geometric factor, we can make an estimate of the best-fit value of the differential sky signal $\hat{\mathcal{F}}$, using the log-likelihood of the form \footnote{The quantities with a hat ($\hat{X}$) denotes the estimator of $X$ from the data.}  
  \begin{eqnarray}\label{es6}
\begin{split}
-2\mathcal{L}_{\mathcal{F}}\propto&\sum_t \sum_i\bigg(\Delta \hat \Omega_{\mathcal{F}}(\hat n_i, f, t)-   \gamma(\hat n_i, f, t)\mathcal{F}(\hat n_i, f, t)\bigg)^\dagger  \\& N_\mathcal{F}^{-1}\bigg(\Delta \hat \Omega_{\mathcal{F}}(\hat n_i, f, t)-   \gamma(\hat n_i, f, t)\mathcal{F}(\hat n_i, f, t)\bigg),
\end{split}
 \end{eqnarray}
where the sum over indices $i$ and $t$ runs over the number of pixels ($N_{pix}$) and total number of time-bins $T_{bin}= T_{obs}/ \Delta t$,  $\mathcal{F}(\hat n, f, t)$ denotes the model of the spectral-derivative signal given in Eq. \ref{sgwbft2}, and $N_\mathcal{F}$ denotes the covariance matrix given by  
  \begin{eqnarray}\label{es6cov}
N_\mathcal{F}= (\mathcal{N}_i(f)\mathcal{N}_j(f) + \mathcal{N}_i(f')\mathcal{N}_j(f')) \delta_{tt'}\delta_{ff'},
 \end{eqnarray}
where we have assumed that the noise is the dominant contribution than the sky signal and have also assumed that the noise is uncorrelated between different frequency channels, different observation time and different pairs of detectors. As expected, the noise covariance matrix for the derivative signal gets contribution from both the frequency channels $f$ and $f'$, and is going to be more noise dominant in comparison to the measurement of $\Omega^s(f)$.    

\textbf{Temporal-derivative: } The time-derivative signal of the SGWB can be estimated from the data using the form 
\begin{align}\label{es7}
\Delta \hat \Omega_{\mathcal{T}} \equiv\, &  |\Omega^d(f, t)-  \Omega^d(f, t+\Delta t)|, \nonumber \\=& \int d^2 \hat n \gamma(\hat n, f, t)|(\Omega^s(\hat n, f, t)- \Omega^s(\hat n, f, t+\Delta t))|,
 \end{align}
 where $\Delta t$ is a small time interval, over which the detector response function has not changed significantly. The corresponding likelihood to obtain the best-fit value can be written as
 \begin{align}\label{es8}
\begin{split}
-2\mathcal{L}_{\mathcal{T}}&\propto  \sum_t \sum_i\bigg(\Delta \hat \Omega_{\mathcal{T}}(\hat n_i, f, t)-  \gamma(\hat n, f, t)\mathcal{T}(\hat n, f, t, t')\bigg)^\dagger \\ & N_\mathcal{T}^{-1} \bigg(\Delta \hat \Omega_{\mathcal{T}}(\hat n_i, f, t,t')-   \gamma(\hat n, f, t)\mathcal{T}(\hat n, f, t,t')\bigg),
\end{split}
 \end{align}
where the index $t$ is summed over the number of temporal bins $T_{bins}= T_{obs}/\Delta t$, index $i$ runs over number of sky pixels and $\mathcal{N}_\mathcal{T}$ denotes the covariance matrix
  \begin{align}\label{es8cov}
N_\mathcal{T}= 2\mathcal{N}_i(f)\mathcal{N}_j(f) \delta_{tt'}\delta_{ff'},
 \end{align}  
where we have assumed again that the noise between two detectors are uncorrelated in time and frequency.
 
\textbf{Tomographic redshift estimate:} In order to separate the cosmological and astrophysical signal as discussed in Sec. \ref{position}, we work in the spherical harmonics space, and define $\Omega_{\mathcal{X}}(\hat n, f, t)$ for  $\mathcal{X}\in \mathcal{F}, \, \mathcal{T}$ and  cosmic density field $\delta(\hat n, z)$ as
 \begin{eqnarray}\label{es9}
\Delta \Omega_{lm|\mathcal{X}}( f, t)= \int d^2\hat n\, \Omega_\mathcal{X}(\hat n, f, t) Y^*_{lm} (\hat n), \nonumber \\ 
\delta_{lm}(z)= \int d^2\hat n\, \delta(\hat n, z) Y^*_{lm} (\hat n).
 \end{eqnarray}
The maximum value of $l$ is related to the smallest angular scale which can be resolved from the experiment. For the GW  interferometer detectors at a distance $D$ apart from each other, the smallest angular scale is diffraction-limited, and this sets the maximum value of $l$ as $l_{max}= 2\pi f D/c$ (where $c$ is the speed of light). For a higher signal to noise ratio (snr), we need to go to high $l$ values, hence large spatial separation between the pair of detectors is required. 

Using $\hat{\mathcal{T}}(\hat n, f, t, \Delta t)$ and $\hat{\mathcal{F}}(\hat n, f, t)$ in the spherical harmonic basis, we can define the cross-correlation with the cosmic density field as
  \begin{eqnarray}\label{es10}
\hat{C}_l^{\mathcal{F}\delta}(f, t, z)= \sum_m\frac{\hat{\mathcal{F}}_{lm}(f, t)\delta^*_{lm}(z)}{2l+1},\nonumber \\
\hat{C}_l^{\mathcal{T}\delta}(f, t, \Delta t, z)= \sum_m\frac{\hat{\mathcal{T}}_{lm}(f, t, \Delta t)\delta^*_{lm}(z)}{2l+1}.
 \end{eqnarray}
Using the cross-correlation power spectrum of the SGWB data with the galaxy surveys, we can estimate the best-fit astrophysical event-rate by using the likelihood   
 \begin{align}\label{es11}
-2\mathcal{L}\propto& \sum_t \sum_l\bigg[\bigg(\hat{C}_l^{\mathcal{X}\delta}(f, t, z)- C_l^{\mathcal{X}\delta}(f, t, z)\bigg)^\dagger N_{{ll'|\mathcal{X}\delta}}^{-1}\nonumber \\&\bigg(\hat{C}_{l'}^{\mathcal{X}\delta}(f, t, z)- C_{l'}^{\mathcal{X}\delta}(f, t, z)\bigg)\bigg],
 \end{align}
 where, $N_{{ll'|\mathcal{X}\delta}}$ is the covariance matrix which can be calculated from the GW detector noise, network of the GW detectors (which affect the value of $l_{max}$), shot noise of the galaxy surveys, and the sky fraction of the galaxy surveys. $C_l^{\mathcal{X}\delta}$ denotes the model of the cross-correlation power spectrum given in  Eq. \ref{sgwbfco1} and in Eq. \ref{sgwbfco2}, where the time-dependence arises only from the bias terms $b_\mathcal{F}$ and $b_{\mathcal{T}}$. However, the power spectrum of the dark matter distribution $P_{DM}(k)$ remains constant in time. As a result, a best-fit value of the bias parameters $b_\mathcal{F}$ and $b_{\mathcal{T}}$ and their time dependence can be inferred by minimizing Eq. \ref{es11}. Though this remains  an interesting avenue to better understand the redshift dependence of the bias, its applicability will depend on the angular resolution $\Delta \theta$, the overlap reduction function $\gamma (\hat n, f, t)$, and the detector noise for different configurations of the future GW detectors network. With the currently available network of detectors LIGO Hanford (H), LIGO Livingston (L) and Virgo (V), we are not going to achieve the required sensitivity to measure the redshift dependent bias of the GW sources from the data.
      
 \section{Error estimation for measuring temporal fluctuations in SGWB}\label{error-estimate}
\begin{figure}
    \centering
    \includegraphics[trim={0 0 0 0cm}, clip, width=0.5\textwidth]{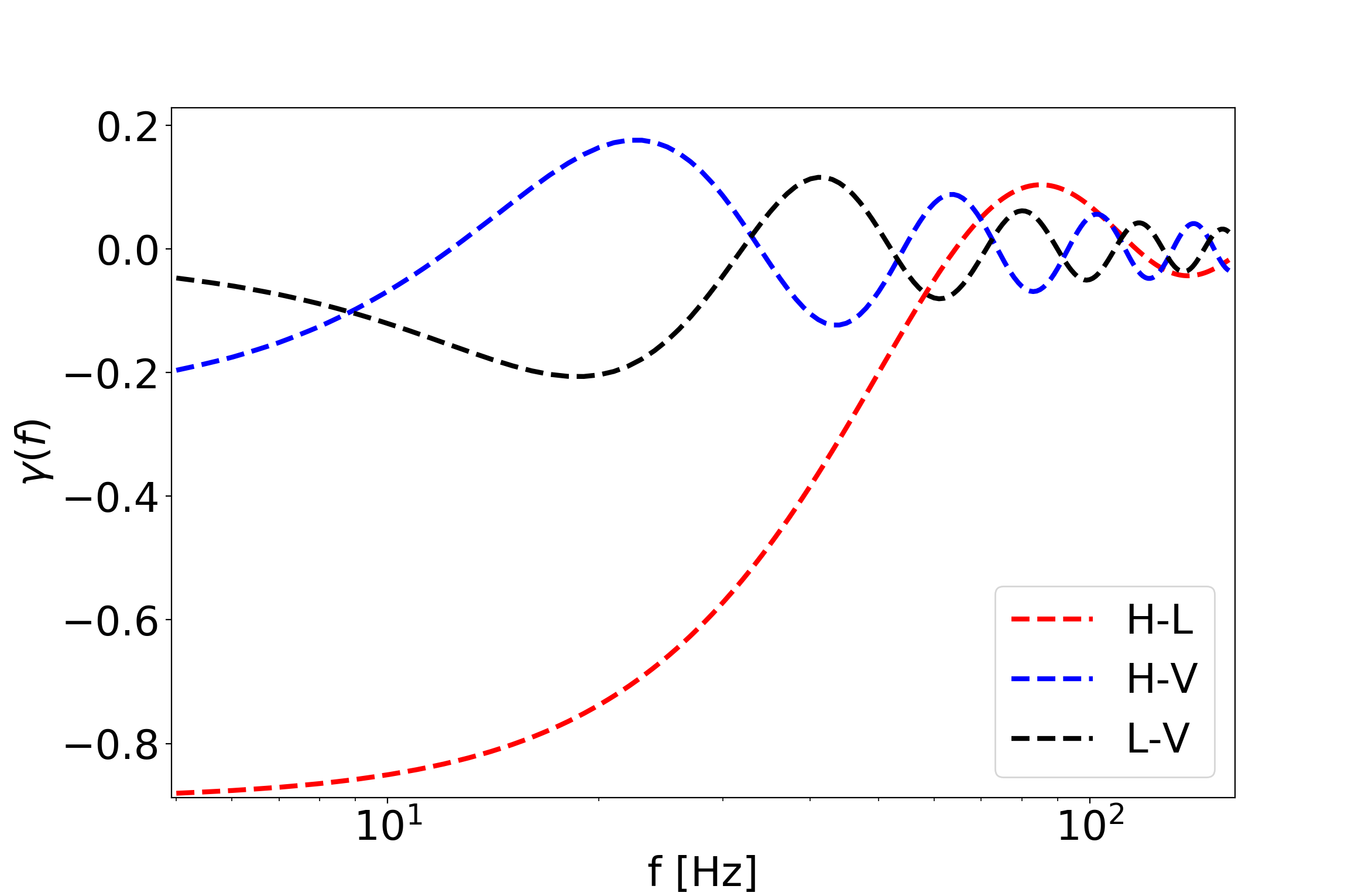}
    \caption{The normalized overlap reduction function as a function of frequency $f$ are shown for the three pairs of detectors LIGO Hanford- LIGO Livingston (H-L), LIGO Hanford- Virgo (H-V), and LIGO Livingston-Virgo (L-V).}
    \label{fig:gamma}
\end{figure}

 We make a Fisher analysis to estimate the error-bar using the Cramer-Rao bound ($\sigma_{\Delta \Omega}\geq \sqrt{\mathbf{F}^{-1}}$) of the time-variability and frequency-variability of the SGWB signal which can be measured using network of GW detectors such as advanced-LIGO Hanford (H), advanced-LIGO Livingston (L), advanced Virgo (V)  {and Cosmic Explorer (CE)}. The rms fluctuation in the SGWB can be expressed as the modification in the amplitude given by $\Delta \Omega (f)= \sqrt{\bigg\langle \bigg(\frac{\Omega_{GW} (t,f) - \bar \Omega_{GW} (f)}{\bar \Omega_{GW} (f)}\bigg)^2\bigg\rangle_T}$. The corresponding Fisher estimate for a network of detectors (I,J) can be written as
 \begin{align}
     \begin{split}
         F_{\alpha\beta}= \int_0^T dt & \sum_I^N\sum_{J>I}^N\bigg(\frac{3H_0^2}{2\pi^2}\bigg)^2 \\ & \int_0^\infty df \frac{\gamma^2(f)\partial_\alpha\Omega_{GW}(f)\partial_\beta\Omega_{GW}(f)}{f^6 \mathcal{N}_I(f)\mathcal{N}_J(f)}, 
     \end{split}
 \end{align}
 where, $\gamma(f)$ is the unnormalized overlap reduction function which depends upon the location and orientation of LIGO-H, LIGO-L and Virgo \footnote{The normalized overlap reduction function is obtained from this file  \href{https://dcc.ligo.org/public/0022/P1000128/026/figure1.dat}{P1000128/026}.}. The overlap function for LIGO-L and LIGO-H (L-H), LIGO-L and Virgo (L-V) and LIGO-H and Virgo (H-V) are shown in the Fig. \ref{fig:gamma}. The first zero crossing of the overlap function takes place at a frequency $f_{char}= c/2D$, where $D$ is the distance between the two detectors \citep{PhysRevD.48.2389}. The $f_{char}$ is smallest for LIGO-H and Virgo, followed by LIGO-L and Virgo and then for LIGO-H and LIGO-L pair of detectors. Only for the frequency range of the GW signal less than $ f \leq f_{char}$, we can make a coincidence detections of the GW signal between a pair of detectors.  As a result,  most of the statistical power for the measurement of the SGWB signal comes from the range of frequency $f \leq f_{char}$. This implies, the most statistical power comes from LIGO-H and LIGO-L pair of detectors, followed by LIGO-L and Virgo and then LIGO-H and Virgo. 
 
\begin{figure}
    \centering
    \includegraphics[trim={0.5cm 0 0.5cm 0.5cm}, clip, width=0.5\textwidth]{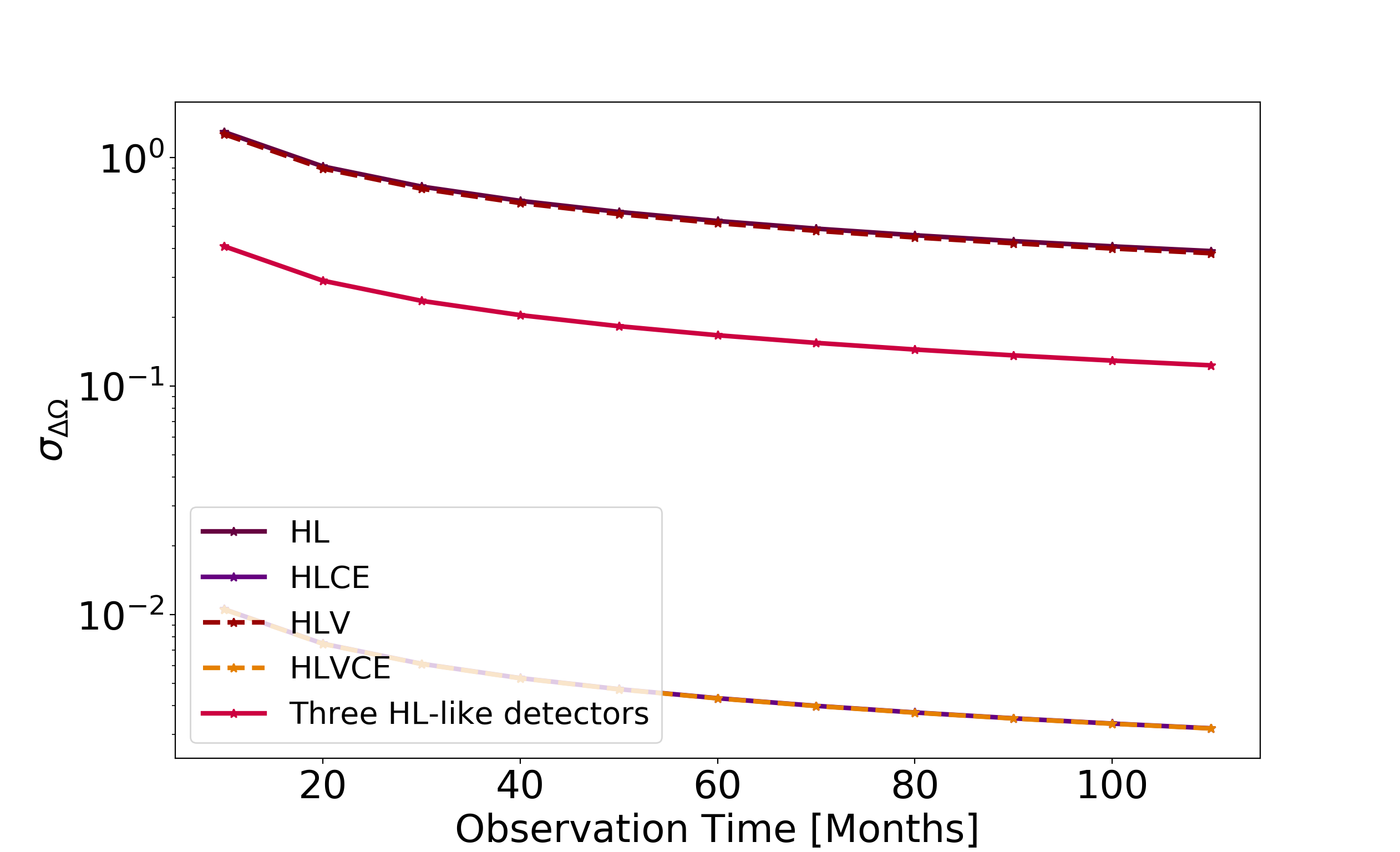}
    \caption{The $1$-$\sigma$ error-bar on the fluctuations in the amplitude of the GW signal for different configurations of the GW detectors LIGO-Hanford (H), LIGO-Livingston (L) and Virgo (V).  {For the case of a three HL-like detector network, we have assumed a configuration with the same detector noise and overlap reduction function as LIGO-H and LIGO-L. For the next generation GW detectors such as Cosmic Explorer (CE) we have considered the instrument noise according to \citet{{Evans:2016mbw}} and with the overlap function same as for the LIGO-H and LIGO-L detector pair.}}
    \label{fig:error}
\end{figure}

 Using the instrument noise for advanced-LIGO and VIRGO,  we obtain the $1$-$\sigma$ error bar on $\Delta \Omega$ in Fig. \ref{fig:error} as a function of the observation time. For the network of three detectors such as LIGO-H, LIGO-L and Virgo, most of the statistical power in the measurement of the signal comes from LIGO-H and LIGO-L for the frequency of GW SGWB signal $f> 30$ Hz.  For a network of three HL-like detectors, we have assumed the same noise curves and the overlap function $\gamma (f)$  as LIGO-H and LIGO-L, and show the amount of gain possible from such a configuration with detector noise of current detectors. 
 
 {For the next generation GW detectors such as Cosmic Explorer (CE) \citep{Evans:2016mbw}, we have assumed the same overlap function as between LIGO-H and LIGO-L, and obtain the $1$-$\sigma$ error-bar on $\Delta \Omega$ of the signal for the combination of three (LIGO-H, LIGO-L, and CE) and four (LIGO-H, LIGO-L, Virgo, and CE) network of detectors. As expected, the gain in the measurability of the signal improves by more than an order of magnitude than with the current ongoing GW experiments. As a result, we can measure the fluctuations in the SGWB with high snr from the future experiments. The difference in the $\sigma_{\Delta \Omega}$ are negligible for the case with or without including Virgo are due to two reasons, (i) the fact that the $f_{char}$ is small than for the case with LIGO-H and LIGO-L, and (ii) the detector noise for Virgo interferometer is higher than CE.}

\section{Conclusion}\label{conclusion}
The SGWB is a rich source of cosmological and astrophysical informations.  {The astrophysical SGWB arises from a large number of coalescing binary systems of compact objects such as black hole, neutron star and white dwarf. Such events are expected to occur randomly following a Poisson distributed in time with a constant event rate which depends upon the astrophysical parameters such as the mass of the compact objects, star formation history, redshift of the source, etc. Due to the Poissonian nature of the GW events, the SGWB is going to exhibit a temporal fluctuations due to the variation of the number of events and is expected to  attain the time translation symmetry on averaging over a long observation time. We point out the temporal dependence of the SGWB in this paper and discuss the implications of measuring it from the ongoing and future GW experiments.} We show that the time-dependence of the SGWB and its rms fluctuations can be a useful probe to learn about several aspects such as the event rate of the GW sources for different chirp masses of the coalescing binaries, the duty cycle of GW signals, and the redshift distribution of the event rate. 

 {The temporal fluctuations of the SGWB can be studied in frequency domain, temporal domain and in the spatial domain as discussed in Sec. \ref{aspects}. By constructing quantities such as the spectral-derivative of the SGWB  denoted by $\mathcal{F}$ we can estimate the redshift integrated total GW energy density arising from different mass windows of the compact objects. The time-derivative of the SGWB denoted by $\mathcal{T}$ captures the temporal fluctuation in the SGWB signal which is related to the event rate of the GW signal contributing at a particular frequency of the SGWB. The spatial distribution of the astrophysical GW sources are expected to follow the spatial distribution of the galaxies and can be estimated by cross-correlating with the galaxy distribution available from the upcoming cosmological surveys (such as DESI \citep{Aghamousa:2016zmz}, EUCLID \citep{2010arXiv1001.0061R}, LSST \citep{2009arXiv0912.0201L}, SPHEREx \citep{Dore:2018kgp}, WFIRST \citep{Dore:2018smn}). The cross-correlation of the SGWB with the galaxy field can probe the time-dependent bias of the SGWB signal for different cosmological redshifts as we have discussed in Sec. \ref{formalism}. The statistical significance of the cross-correlation between the SGWB map with the galaxy distribution is going to be limited by the angular resolution of the SGWB map. For the interferometer-based GW detectors, the smallest angular scale $\Delta \theta$ that can be resolved in SGWB sky is set by the diffraction limit, which can be written in terms of the frequency of the GW signal and the distance between two detectors $D$ as $\Delta \theta= c/2f D$. This implies that the interferometer detectors, which are farthest away are the best for making high-resolution maps of the SGWB signal. However, for the interferometer detectors which are farther away are going to have an overlap reduction function with smaller value of $f_{char}= c/2D$, which can result into a higher instrument noise for frequencies greater than $f_{char}$. } 

We propose that  data analysis  performed for ongoing GW experiments should search for temporal fluctuations along with the algorithms which search for spatial fluctuations. This procedure will be potentially useful for identifying the event rates for different binary species such as BNS, BBHs and BH-NS systems. The procedure  outlined here will help establish the time-translation symmetry of the SGWB and inform us  about the astrophysics related to the formation of these binary sources. This avenue of research will provide a useful tool for  distinguishing between the astrophysical and cosmological SGWB signals. By exploring the time-dependence aspect of the SGWB signal, we can remove contamination from the astrophysical SGWB signal and  peer into the higher frequency $f \in 10-1000$ Hz cosmological SGWB signals originating at different epochs of the Universe \citep{Starobinsky:1979ty,Turner:1996ck, Martin:2013nzq,Kibble:1976sj,Damour:2004kw,Kosowsky:1992rz,Kamionkowski:1993fg}. 
   
\section*{Acknowledgement}
The results of this analysis are obtained using the Horizon cluster hosted by Institut d'Astrophysique de Paris. We thank Stephane Rouberol for smoothly running the Horizon cluster. The work of S.M. is supported by the Labex ILP (reference ANR-10-LABX-63) part of the Idex SUPER, and received financial state aid managed by the Agence Nationale de la Recherche, as part of the programme Investissements d'avenir under the reference ANR-11-IDEX-0004-02.

\bibliographystyle{mnras}
\bibliography{SGWB-Astrophysics-3}
\label{lastpage}

\end{document}